\documentclass[jcp,twocolumn,showpacs,floats]{revtex4}
\usepackage{graphicx}
\usepackage{amsmath}
\usepackage{tikz}
\usepackage{color}
\graphicspath{{images/},{images/pdf/}}
\begin{document}
\title{Chiral selectivity of amino acid adsorption on chiral surfaces - the case of alanine on Pt}
\author{J.-H.~Franke} 
\author{D.S.~Kosov}
\altaffiliation{College of Science, Technology and Engineering, James Cook University,Townsville, QLD, 4811, Australia}
\affiliation{Department of Physics, Campus Plaine - CP 231, Universite Libre de Bruxelles, 1050 Brussels, Belgium}
\date{\today}

\begin{abstract}
We study the binding pattern of the amino acid alanine on the naturally chiral Pt surfaces Pt(531), Pt(321) and Pt(643). These surfaces are all vicinal to the \{111\} direction but have different local environments of their kink sites and are thus a model for realistic roughened Pt surfaces. Alanine has only a single methyl group attached to its chiral center, which makes the number of possible binding conformations computationally tractable. Additionally, only the amine and carboxyl group are expected to interact strongly with the Pt substrate. On Pt(531) we study the molecule in its pristine as well as its deprotonated form and find that the deprotonated one is more stable by 0.39 eV. Therefore, we study the molecule in its deprotonated form on Pt(321) and Pt(643). As expected, the oxygen and nitrogen atoms of the deprotonated molecule provide a local binding "tripod" and the most stable adsorption configurations optimize the interaction of this "tripod" with undercoordinated surface atoms. However, the interaction of the methyl group plays an important role: it induces significant chiral selectivity of about 60 meV on all surfaces. Hereby, the L-enantiomer adsorbs preferentially to the Pt(321)$^S$ and Pt(643)$^S$ surfaces while the D-enantiomer is more stable on Pt(531)$^S$. The binding energies increase with increasing surface density of kink sites, i.e. they are largest for Pt(531)$^S$ and smallest for Pt(643)$^S$.
\end{abstract}
\maketitle

\begin{figure*}[Htb]
\includegraphics[width=12cm]{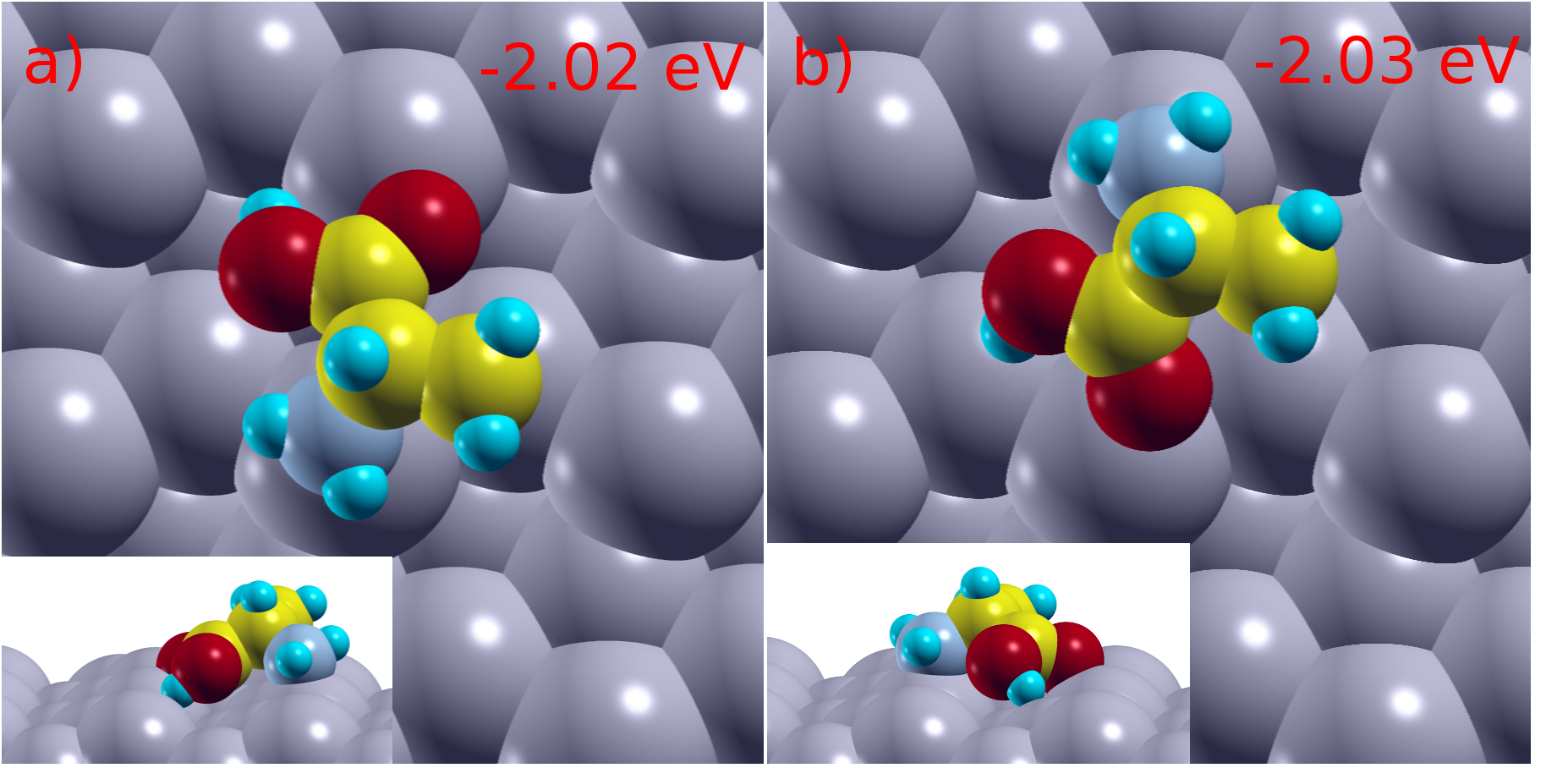}
\caption{The most stable configurations of pristine L-alanine (\textbf{a}) and D-alanine (\textbf{b}) on Pt(531)$^S$. }
\label{fig:geom531_prist}
\end{figure*}

\section{Introduction}

Almost all biological molecules (proteins, DNA, RNA, sugars and peptides) cannot be superimposed with their mirror image -- they are chiral. From general  physical consideration one expects that left- and right-handed molecules should play the same role in biochemistry, and for some unknown and hotly debated reason our life is homochiral.\cite{Blackmond2010} It has been suggested, that homochirality of chiral molecules in the prebiotic environment was a result of symmetry breaking by the absorption on naturally chiral mineral surfaces.\cite{Hazen2001,Hazen2010} The two mirror images, called enantiomers, interact differently with other chiral molecules as a result of their different geometry.\cite{Gellman2010} Consequently, the control over molecular chirality is important for drug molecules that are designed to interact only with certain biomolecules while minimizing interactions with others to avoid unwanted side effects.\cite{Rekoske2001} This has sparked a huge interest in studies of chiral systems to understand the basics of chirally specific interactions that can then be employed to produce enantiopure molecules. Different possibilities exist towards this end, the most industrially relevant being the synthesis via homogenous catalyis.\cite{Blaser2005} Other possibilities are the separation of the enantiomers after synthesis or the use of heterogeneous catalysis.\cite{Rekoske2001,Mallat2007} The latter is mostly done on surfaces rendered chiral by chiral adsorbates.\cite{Mallat2007,Kyriakou2011,Lawton2013,Meemken2012,Gross2013,Schmidt2012,Lorenzo2000,Fasel2006} However, also simple metal crystals can be intrinsically chiral when they are cut in certain ways.\cite{Sholl2009,Clegg2010} The interaction of chiral molecules with these surfaces can be studied by surface science and computational methods to gain a fundamental understanding of enantioselectivity.\cite{Sholl1998,Sholl2001,Held2008,Clegg2011,Ahmadi1999,Horvath2004,Bombis2010,Bhatia2005,Bhatia2008,Huang2011,Huang2008,James2008,Sljivancanin2002,Vidal2005,Yuk2014,Yun2013} 

Among the most widely studied chiral systems are simple amino acids on naturally chiral metal surfaces.\cite{Clegg2011,Han2012,Thomsen2012,Greber2006,James2008,Yun2013,Yun2014,Sljivancanin2002} A lot of experimental and computational studies were dedicated to amino acids on Cu surfaces,\cite{Clegg2011,Zhao1999,Thomsen2012,Han2012,Yun2013,Yun2014,Jones2006,Eralp2010,Eralp2011,Cheong2011,Sayago2005,Gladys2007,James2008,James2008a,Rankin2005,Han2011} but also on Pd, Pt and Au surfaces.\cite{Mahapatra2014,James2014,James2008,James2008a,Gao2007,Han2011,Kuhnle2002} On Cu it was found that the carboxyl group of amino acids is deprotonated and the molecule adsorbs on most surfaces in a "tridentate" fashion, i.e. with the two oxygen atoms of the carboxyl group and the nitrogen atom from the amine group binding to the substrate. For alaninate on Cu(531) this binding tripod adsorbs on \{311\} and \{110\} microfacets of the surface with different probabilities for each binding pattern for the two enantiomers and enantioselective adsorption energies.\cite{Gladys2007} On Cu(421) enantioselectivities are smaller, but adsorption geometries and decomposition products are chirally selective.\cite{Thomsen2012} On Cu(643) Density Functional Theory (DFT) calculations show that alaninate binds with the nitrogen group to kink atoms and with the carboxyl group to the corresponding ridge and no significant enantioselectivity is found.\cite{Han2012} On Cu(3,1,17) experiments and calculations show again no enantioselectivity.\cite{Yun2014} On Au surfaces, mostly cysteine was studied that interacts strongly through its sulfur containing side-group and exhibits significant enantioselectivity on Au(17 11 9).\cite{Greber2006} 

On Pd surfaces, experiments indicate that alanine adsorbs in its zwitterionic and deprotonated form while DFT calculations indicate that deprotonated molecules are more stable.\cite{James2008a,Mahapatra2014} The difference between calculations and experiment can be explained by the neglected influence of molecule-molecule interactions in the calculations that stabilize the zwitterionic form.\cite{Mahapatra2014,Han2011} On Pt, experiments also show the presence of zwitterionic molecules, while DFT calculations show that the pristine state of the similar amino acid glycine would be the most stable, 0.24 eV more stable than the deprotonated form.\cite{Han2011} In line with these calculations, recent experiments show that glycine adsorbs in its pristine form on Pt(111) at low coverages, showing that, similar to Pd(111) surfaces the molecule-molecule interaction stabilizes the zwitterionic form.\cite{Shavorskiy2013,Han2011} In the present paper, we concern ourselves with low-coverage adsorption where molecule-molecule interaction is negligible. 

Herein we present calculations of alanine on naturally chiral Pt surfaces vicinal to the \{111\} direction, Pt(531), Pt(321) and Pt(643). On Pt(531) we study the molecule in its pristine and deprotonated form and establish that the deprotonated form is more stable by 0.39 eV, in contrast to glycine on Pt(111) where the pristine molecule was found to be more stable.\cite{Han2011}. Entropy gains by desorption of molecular hydrogen should further favor the adsorption of deprotonated molecules.\cite{Bebensee2013} Therefore we continue to study the molecule in its deprotonated form on Pt(321) and Pt(643). We find that the binding energy of the deprotonated molecule is larger on Pt(531) then on Pt(321) and Pt(643). The calculated chiral selectivity amounts to about 60 meV for all surfaces with a preference of D-alaninate on the S surfaces of Pt(321) and Pt(643). Contrary to these results, L-alaninate is found to be more stable on Pt(531)$^S$. The trend in binding energy can be understood by the binding motives exhibited by the alaninate molecule on the studied surfaces. In its most stable configurations, the molecule adsorbs with the two oxygen atoms of the carboxyl group and the nitrogen atom of the amine group to undercoordinated Pt atoms. On Pt(643) only one of the three binding sites can bind to a kink atom, with the other two binding to ridge atoms. On Pt(321) and Pt(531) two of the three atoms bind to kink atoms. On Pt(531) the (111) terraces are sufficiently small to be bridged by the molecule, which better accomodates the methyl group and thus leads to higher binding energies. When looking at the two most stable configurations for both enantiomers, an interesting pattern arises: all 4 configurations can be constructed by two binding patterns for the oxygen and nitrogen atoms and the exchange of the methyl group with the hydrogen atom on the chiral center. Thus, there is a competition between an optimal positioning of the methyl group and the optimal position for the binding tripod. On Pt(531) and Pt(321), the positioning of the binding tripod dominates as it is the same for the most stable configurations, while on Pt(643) the positioning of the methyl group is more important.

The paper is organized as follows. In section II we outline the computational method used in our simulations. In section III we describe the results obtained for alanine and alaninate on Pt(531), while the results for alaninate on Pt(321) and Pt(643) are presented in Section IV and V, respectively. Section VI concludes the paper. 
\begin{figure*}[Htb]
\includegraphics[width=12cm]{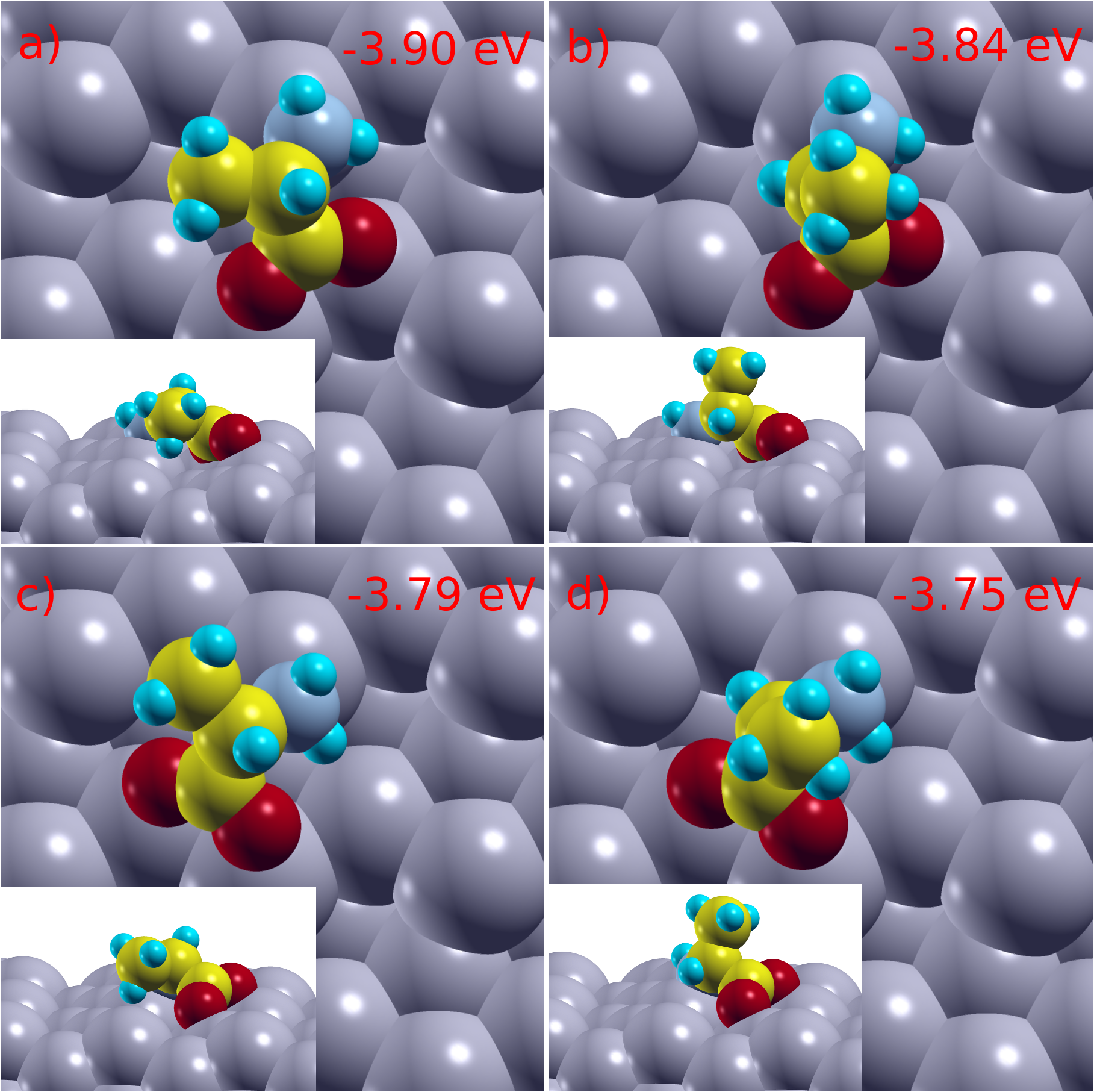}
\caption{Alaninate adsorbed on Pt(531)$^S$. The most stable (\textbf{a},\textbf{b}) and second most stable configurations (\textbf{c},\textbf{d}) are given for L-alaninate (\textbf{a},\textbf{c}) and D-alaninate (\textbf{b},\textbf{d}). }
\label{fig:geom531}
\end{figure*}

\begin{figure*}[Htb]
\includegraphics[width=12cm]{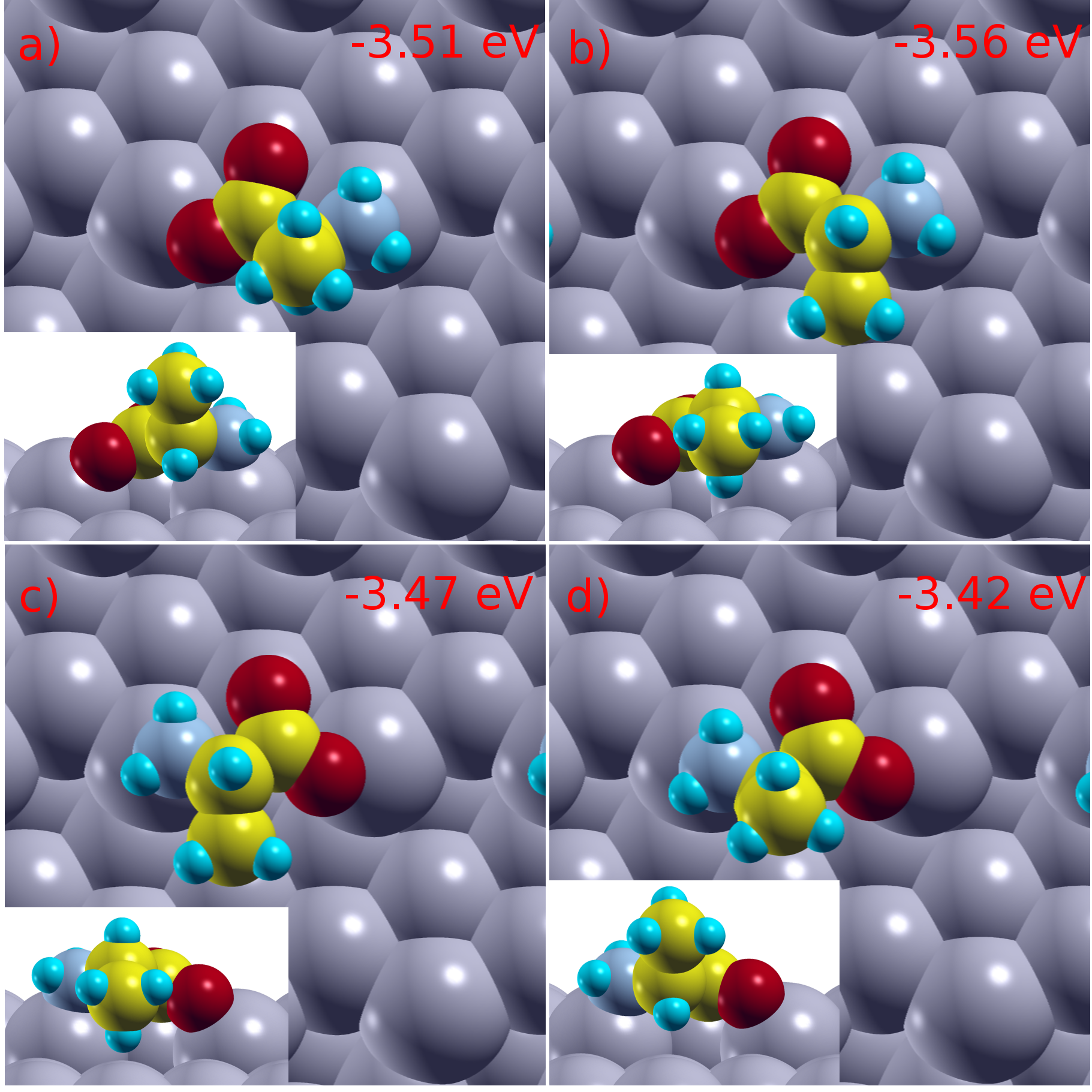}
\caption{Alaninate adsorbed on Pt(321)$^S$. The most stable (\textbf{a},\textbf{b}) and second most stable configurations (\textbf{c},\textbf{d}) are given for L-alaninate (\textbf{a},\textbf{c}) and D-alaninate (\textbf{b},\textbf{d}).}
\label{fig:geom321}
\end{figure*}

\begin{figure*}[Htb]
\includegraphics[width=12cm]{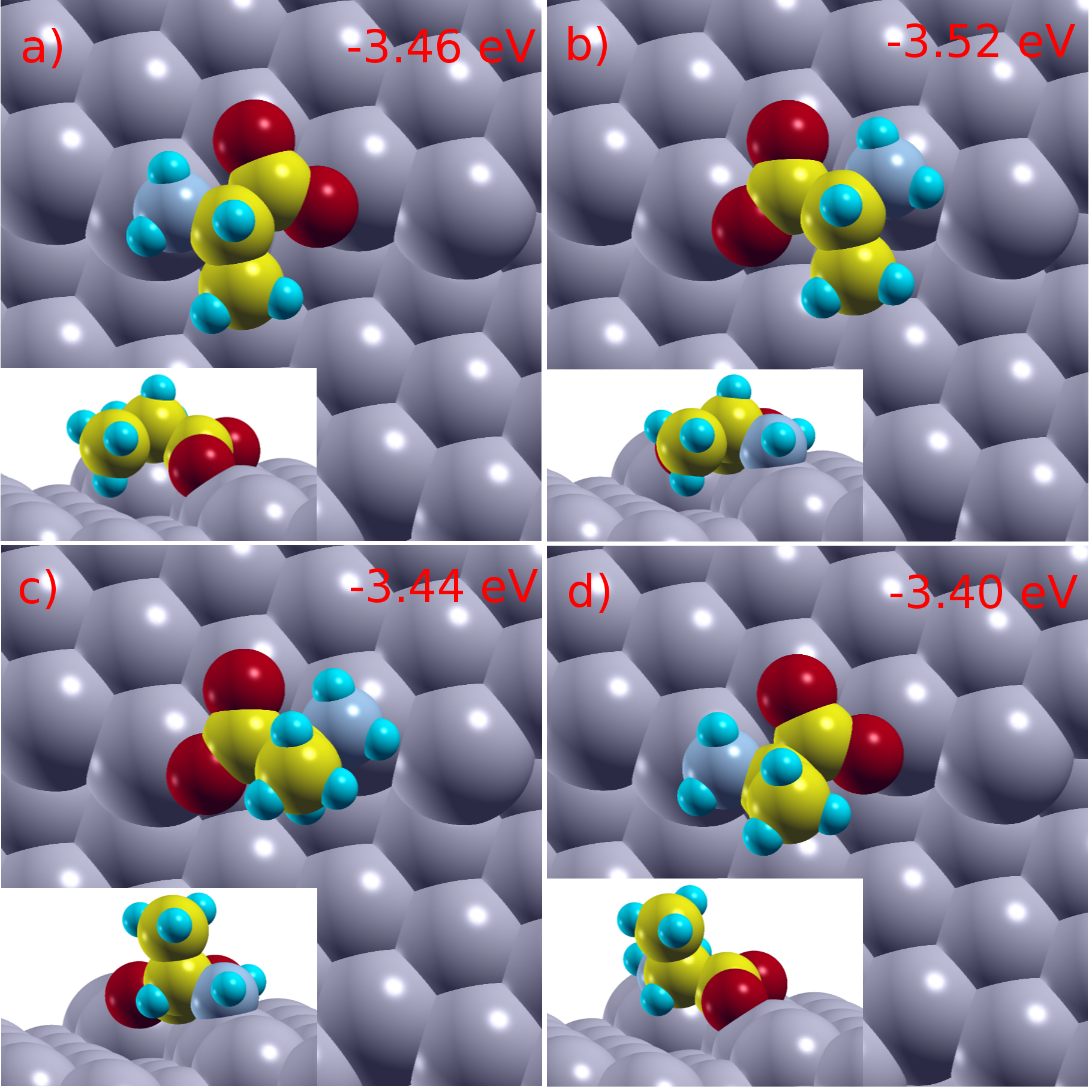}
\caption{Alaninate adsorbed on Pt(643)$^S$. The most stable (\textbf{a},\textbf{b}) and second most stable configurations (\textbf{c},\textbf{d}) are given for L-alaninate (\textbf{a},\textbf{c}) and D-alaninate (\textbf{b},\textbf{d}).}
\label{fig:geom643}
\end{figure*}

\section{Computational details}

We obtained our results with the DFT code VASP 5.3.2\cite{Hafner2008,Kresse1996a,Kresse1996b} using the oPBE-vdW functional\cite{Klimes2010,Klimes2011}. We opted for this functional as van der Waals interactions were found to be important for the correct description of weakly bound molecules on surfaces.\cite{Franke2013,Franke2013a,Wei2014} The Projector Augmented Wave method\cite{Bloechl1994,Kresse1999} was employed with an energy cutoff of 400 eV. Structural relaxations were stopped when all forces were smaller than 10 meV/\AA\ with wavefunctions converged to energy changes between successive steps smaller than 10$^{-5}$ eV. For all slab calculations dipole corrections to the potential are applied throughout.\cite{Neugebauer1992} 

The (643), (321) and (531) surfaces were constructed with a thickness of 56, 28 and 16 layers of (643), (321) and (531) orientation, respectively. The molecule was adsorbed on (2x1), (2x2) and (3x3) supercells of the (643), (321) and (531) surfaces on one side of the slab that corresponds to the S chirality of the surface.\cite{Ahmadi1999,Sholl2001} For all surfaces the upper half of the slab was allowed to relax as were all atoms of the molecule. The K-mesh employed was 3x3x1 for all surfaces and a Gaussian broadening of 0.1 eV was used to facilitate convergence. Adsorption energies $E_{adsorption}$ are given with reference to the isolated surface $E_{surface}$ relaxed upon removing the molecule from the unit cell using identical computational parameters and the energy of the molecule $E_{molecule}$

\begin{equation}
E_{adsorption} = E_{mol\ on\ surface} - E_{surface} - E_{molecule}.
\end{equation}

To compare the energies of the adsorbed pristine molecule with its deprotonated form on Pt(531), we introduce an additional hydrogen atom on a kink site away from the deprotonated molecule. The kink site was found to be the most stable adsorption site of hydrogen on Pt(531). 

To better understand the origins of the observed binding patterns we calculate the contributions to the binding energy of each chemical group attached to the chiral center.\cite{Sljivancanin2002} To this end we cut away all other components of the molecule, freeze all atoms and saturate the dangling bond with an additional hydrogen atom that is free to relax. Moreover, we calculate the energy of this frozen molecular species in vacuum with the extra hydrogen again free to relax and the frozen substrate separately. The adsorption energy of the molecular fragment is then calculated as

\begin{align}
E_{adsorption\ of\ fragment} =& \ E_{fragment\ on\ surface} \nonumber \\
                              &- E_{frozen\ surface} \nonumber \\
                              &- E_{frozen\ fragment}.
\end{align}

Lastly, we calculate the deformation energy of the substrate and the molecule as the energy difference between its frozen and relaxed state.

\begin{table*}[Htb]
\caption{Adsorption energy component analysis for all alaninate surface configurations. Adsorption energies of hydrogen saturated CH3, NH2 and COO groups in the frozen adsorption geometries and deformation energies of the substrate and molecule are calculated. The sum of the binding energy components considered is comparable to the total binding energy, showing that the binding energy decomposition is roughly correct in the systems at hand (cf. Fig.  \ref{fig:geom531} - \ref{fig:geom643}). The most stable and second most stable configurations of the L-enantiomer are labelled L1 and L2, respectively.}
\begin{tabular}{c|c|c|c|c|c|c}
\hline
\multicolumn{1}{c}{alaninate on} & \multicolumn{2}{|c}{deformation energy (eV) } & \multicolumn{3}{|c}{adsorption energy (eV)} & \multicolumn{1}{|c}{sum of components (eV)} \\
 & surface & molecule & CH3 group & NH2 group & COO group &    \\
\hline
L1 on Pt(531)$^S$    & 0.15 & 0.65 & -0.29 (-0.40) & -1.60 (-0.69) & -3.53 (-1.07) & -4.61 \\
L2 on Pt(531)$^S$    & 0.22 & 0.68 & -0.19 (-0.34) & -1.51 (-0.65) & -3.49 (-1.04) & -4.30 \\
D1 on Pt(531)$^S$    & 0.15 & 0.67 & -0.19 (-0.17) & -1.60 (-0.68) & -3.50 (-1.07) & -4.48 \\
D2 on Pt(531)$^S$    & 0.22 & 0.69 & -0.13 (-0.13) & -1.52 (-0.65) & -3.47 (-1.05) & -4.20 \\

L1 on Pt(321)$^S$    & 0.16 & 0.48 & -0.13 (-0.14) & -1.36 (-0.61) & -2.97 (-0.99) & -3.82 \\
L2 on Pt(321)$^S$    & 0.24 & 0.49 & -0.25 (-0.41) & -1.39 (-0.61) & -2.87 (-0.35) & -3.78 \\
D1 on Pt(321)$^S$    & 0.16 & 0.49 & -0.22 (-0.39) & -1.35 (-0.61) & -3.00 (-0.98) & -3.92 \\
D2 on Pt(321)$^S$    & 0.24 & 0.48 & -0.14 (-0.15) & -1.43 (-0.63) & -2.79 (-0.96) & -3.64 \\

L1 on Pt(643)$^S$    & 0.24 & 0.65 & -0.18 (-0.39) & -1.35 (-0.58) & -3.04 (-0.95) & -3.69 \\
L2 on Pt(643)$^S$    & 0.16 & 0.70 & -0.13 (-0.14) & -1.27 (-0.66) & -3.23 (-0.98) & -3.77 \\
D1 on Pt(643)$^S$    & 0.17 & 0.68 & -0.24 (-0.42) & -1.27 (-0.66) & -3.26 (-0.97) & -3.93 \\
D2 on Pt(643)$^S$    & 0.22 & 0.67 & -0.09 (-0.14) & -1.41 (-0.61) & -2.98 (-0.97) & -3.58 \\
\hline
\end{tabular}
\label{tab:decomp}
\end{table*}

\section{Alanine on Pt(531)}

The (531) surface is similar to the (321) surface in the structure of its step edges, albeit with smaller terrace sizes. The terraces are sufficiently small so that the molecule can bridge them on this surface. The alanine molecule can adsorb in different forms, depending on the position of the proton of the carboxyl group. When bound to one of the two oxygen atoms, the molecule is in its pristine form. However, in our calculations we need to differentiate between the two oxygen atoms the proton might bind to, as well as its direction, pointing either away or towards the other molecular side groups. This gives us four configurations we need to calculate for the pristine molecule. These molecular configurations are then adsorbed with either their amine or carboxyl group above a kink atom and rotated in steps of 30 degrees to find the most stable adsorption configuration. We also calculate some zwitterionic adsorption configurations that we find to be significantly less stable by about 0.4 eV. For all deprotonated (alaninate) adsorption configurations on all surfaces studied, the most stable conformation is found by putting the molecule with either its nitrogen or oxygen atoms over a kink atom and rotating the molecule around this binding site.

First, we identify the most stable configurations of alanine on Pt(531). For both enantiomers, the most stable configurations bridge the terrace and bind with the nitrogen and the  hydrogen-free oxygen atom to kink sites. For both configurations the methyl group is close to the surface (cf. Fig. \ref{fig:geom531_prist}). Binding energies are similar with a slight preference for D-alanine on Pt(531)$^S$ of 12 meV. 

Similar to alanine, terrace bridging conformations are found to be most stable for alaninate on Pt(531). As a result, on Pt(531)$^S$ alaninate is bound to two kink atoms and one ridge atom of neighboring (111) terraces (cf. Fig. \ref{fig:geom531}). The adsorption energies of alaninate are much lower than the adsorption energies of pristine alanine. The largest part of this difference is due to the deprotonation energy, i.e. the energy cost of removing the proton from the pristine molecule and forming H$_2$ in vacuum, which we calculate as 1.92 eV for alanine. The dissociative adsorption of H$_2$ on Pt(531) gives an adsorption energy of -0.43 eV per proton, which reduces the deprotonation energy on the surface to 1.49 eV.

For alaninate on Pt(531), the binding tripod is the same for the most stable configurations of both enantiomers. Correspondingly, the energy differences between configurations with the same binding tripod pattern are smaller (0.06 eV and 0.04 eV) then the differences between configurations with different tripod binding pattern and the same methyl group configuration (0.11 eV and 0.09 eV). The most stable configuration is L-alaninate on Pt(531)$^S$ and the surface exhibits a chiral selectivity of 0.06 eV. The optimized binding of the tripod can be rationalized by the analysis of the binding energy components. The carboxyl and amine groups are bound more strongly by 0.03 eV and 0.08 eV for the most stable configurations when compared to the second most stable ones (cf. Tab. \ref{tab:decomp}). The methyl group adsorption energy is 0.1 eV lower for the D1 configuration than for L1 thus providing the biggest difference between the two configurations binding energy components and leading to the calculated significant chiral selectivity.

\section{Alaninate on Pt(321)}

On Pt(321)$^S$ the terrace size is larger than on Pt(531) so that the molecule can no longer bridge the terrace. The distance between the kink atoms on one ridge is the same so that the molecule can bind to two kink (three-fold undercoordinated) and one singly undercoordinated atom of the same (111) terrace. Accordingly, adsorption configurations that were already found on Pt(531), but were higher in adsorption energy are the most stable on Pt(321)(cf. Fig. \ref{fig:geom321}). 

Here, the energy differences between configurations with the same positioning of the methyl group (Fig. \ref{fig:geom321} (a)-(d) and (b)-(c)) is 0.09 eV. The energy differences between the configurations with the same binding tripod position is 0.05 eV. As a result, the binding tripod position is the same for the most stable configurations of both enantiomers similar to adsorption on Pt(531). The position of the tripod as well as the methyl group is both optimal in the case of D-alaninate, which is 0.04 eV more stable on Pt(321)$^S$ than on Pt(643)$^S$ and 0.34 eV less stable than L-alaninate on Pt(531)$^S$. The (321) surface shows chiral selectivity of 0.05 eV. The sum of the adsorption energies for the carboxyl and amine groups are lower for the most stable configurations (L1 and D1) of both enantiomers on Pt(321) than for the second most stable ones (L2 and D2). The adsorption energies of the methyl group are lower for D1 and L2 than for D2 and L1, which shows that the former configurations are optimized on the surface, corresponding well to the interpretation derived from the total adsorption energies above. Put another way, the difference in interaction energies of the methyl group for the most stable configuration of each enantiomer leads to the calculated chiral selectivity, similar to alaninate on Pt(531).

\section{Alaninate on Pt(643)}

The Pt(643) surface has a longer ridge when compared to Pt(321) that prevents the molecule from binding to two kink atoms at the same time. As a result the molecule is bound to one three-fold, one two-fold and one singly undercoordinated surface atom of the same (111) terrace. The most stable configurations of both enantiomers exhibit a methyl group lying flat on the surface, while it is oriented more along the surface normal for the second most stable configurations. 

The binding tripod has two binding patterns with either the nitrogen or one oxygen atom bound to the kink atom. The geometry with nitrogen bound to the kink atom is less stable, which can be inferred from the comparison of the binding energies of the most stable configurations (cf. Fig. \ref{fig:geom643}(a, b)). The energy differences between configurations with the same positioning of the methyl group is 0.06 eV and 0.04 eV for the most stable and second-most stable configurations, respectively. The energy differences for the same binding tripod amount to 0.06 eV and 0.08 eV. Thus, the positioning of the methyl group is slightly more important than the positioning of the binding tripod on Pt(643). On Pt(643)$^S$, D-alaninate optimizes both interactions at the same time with a chiral selectivity of 0.06 eV. Again, the adsorption energy component analysis supports these findings. The adsorption energy of the methyl group is lower for the configurations L1 and D1 while the carboxyl and amine groups are bound more strongly for L2 and D1, thus making D1 the overall preferred configuration.

\section{Conclusion}

We studied the adsorption of alanine in its deprotonated form on three chiral Pt surfaces vicinal to the {111} direction. On one of the surfaces, Pt(531), we also calculated the adsorption of pristine and zwitterionic molecular species, finding that they are less stable than the deprotonated species. For alaninate we found that the molecule interacts with the different surfaces via the oxygen and nitrogen atoms of the molecule. An additional interaction, important for the chiral selectivity, is then the interaction of the methyl group with the surface. 

On Pt(643) and Pt(321) the molecule is bound to kinks on a single ridge. On Pt(321) the distance between two kink atoms is sufficiently large that the molecule can interact with two kink atoms at the same time, while it interacts with only one on Pt(643). On Pt(531) the surface terraces are small enough so that the molecule can interact simultaneously with kinks on two different facets. The adsorption energies follow the trend in kink-atom density by decreasing slightly from Pt(643) to Pt(321) and then more sharply from Pt(321) to Pt(531).

We find that on all surfaces, the two most stable configurations of each enantiomer show two distinct binding patterns of the oxygen and nitrogen atoms to the substrate. The two most stable binding configurations of the corresponding chiral partner differ by exchanging the hydrogen and methyl group on the chiral center. The two most stable adsorption configurations of both enantiomers thus constitute a combination of one of the two binding patterns of the binding tripod and one or the other binding pattern of the methyl group. Since for each interaction one is more stable than the other, the most favorable combination of both interactions leads to the preference of a certain enantiomer on each surface. On Pt(643)$^S$ and Pt(321)$^S$ this is the D-enantiomer, while on Pt(531)$^S$ it turns out to be the L-enantiomer. 

Our calculations show the intricacies involved in the binding of amino acids on naturally chiral metal surfaces. Binding motifs can vary from one specific surface facet to the next which can also lead to a change in the sign of the chiral selectivity. This is already the case for the relatively simple amino acid alanine and can be traced back to the interaction of the atoms forming the peptide bond, i.e. this effect should be present for any (chiral) amino acid. The multitude of interactions from more complex side groups in amino acids other than alanine may dramatically complicate the picture further. On a real Pt surface, roughening will probably expose a variety of different facets at the same time. Also, the multitude of reactions the molecule might undergo can be expected to depend on the exact facet the molecule interacts with. The different binding configurations on different facets may favor different reactions over competing processes, as their reaction barriers depend on their exact geometry.

\section{Acknowledgements}

This work has been supported by the Francqui Foundation, and Programme d'Actions de Recherche Concertee de la Communaute Francaise, Belgium. Computational ressources have been provided by the Consortium des Équipements de Calcul Intensif (CÉCI), funded by the Fonds de la Recherche Scientifique de Belgique (F.R.S.-FNRS) under Grant No. 2.5020.11.

\bibliography{paper}
\end{document}